\documentclass[prb,superscriptaddress,twocolumn,showpacs,floatfix]{revtex4}
\usepackage{color,graphicx}
\usepackage{bm}
\usepackage{amsmath}
\usepackage{amssymb}

\begin{document}
\title{Singularity in self-energy and composite fermion excitations of interacting electrons }
\author{ Lijun Zhu} 
\affiliation{Theoretical Division and Center for Nonlinear Studies, Los Alamos National Laboratory, Los Alamos, New Mexico 87545, USA}
\affiliation{Department of Physics and Astronomy, University of California, Riverside, CA 92521, USA}
\author{Jian-Xin Zhu}
\affiliation{Theoretical Division and Center for Nonlinear Studies, Los Alamos National Laboratory, Los Alamos, New Mexico 87545, USA}

\begin{abstract}
We study the coherent excitations of a composite fermion operator $f_{i\sigma}(-1)^{n_{i{\bar \sigma}}}$, where $f_{i\sigma}$ is the fermion operator for interacting electrons and $n_{i{\bar \sigma}}$ is the number operator of electrons with the opposite spin. In the two-impurity Anderson model, we show that the excitation of this composite fermion has a finite spectral weight near the Fermi energy in the regime dominated by inter-site spin exchange coupling where the Kondo fixed point is prevented. From scattering off this coherent composite fermion mode, the excitation of the regular fermion $f_{i\sigma}$ develops a pseudogap and its self-energy is singular. Conversely, when the regular fermion develops Kondo resonance in the Kondo resonance regime, the excitation of the composite fermion develops a pseudogap instead. We argue that the composite fermion could develop a Fermi surface but ``hidden'' from charge excitations in lattice generalizations.  
\end{abstract}
\maketitle

\section{Introduction}
\label{sec:intro}

In strongly correlated electron materials, the Coulomb interaction is often larger than the kinetic energy, leading to localization of electrons, or a Mott insulator. It has been argued that being in the vicinity of a Mott metal-insulator transition is responsible for many emergent phenomena, such as magnetic quantum phase transitions and superconductivity in heavy fermion systems,~\cite{KBqcp} iron-based superconductors,~\cite{Si2008,EChia:2010,Shen2012} and cuprate superconductors.~\cite{Tremblay04}  Studies from the dynamical mean-field theory (DMFT) and its extensions provide insights into the Mott transition from the Kondo dynamics in the Anderson impurity (cluster) model as mapped from the lattice model.~\cite{dmft} A metallic state with non-interacting electron behavior, is associated with the strong-coupling Kondo fixed point: hopping processes involving double occupancy at one lattice site are allowed through the formation of Kondo singlets. When the Kondo fixed point is prevented at low energies (also dubbed as ``Kondo-breakdown''), the electrons are therefore localized to form an insulator. However, in the original single-site DMFT implementation of the Hubbard model, the paramagnetic insulator solution carries an extensive entropy from unscreened local moments, and the single-particle excitation has a sharp excitation gap between lower and upper Hubbard bands. When inter-site coupling is incorporated, e.g., in cluster DMFT calculations, the unscreened local moments can reduce entropy by forming intersite spin singlets. A pseudogap-type of spectrum, with finite single-particle excitations between the lower and upper Hubbard bands but suppressed near the Fermi energy, is typically identified.~\cite{Tremblay04}  Furthermore, it has been shown that the pseudogap is associated with singularities in self-energies of interacting electrons.~\cite{Sakai10} These singularities could form a Luttinger surface where the Green's function $G$ changes sign by $G\to 0$, in contrast with the regular Fermi surface where $G$ changes sign by $G\to \pm \infty$. 

In this paper, we study an origin of the singularity in self-energies (and a pseudogap state) associated with the emergent coherent fermionic excitations in terms of a composite fermion operator ${\bar f}_{i\sigma} = f_{i\sigma}(-1)^{n_{i{\bar \sigma}}}$, where  $f_{i\sigma}$ is the regular fermion operator for an interacting electron at a given site $i$ with spin $\sigma$, and $n_{i{\bar \sigma}}$ is the number operator of electrons with the opposite spin.  As we show, to introduce this composite fermion is necessary for interacting electrons, as the single-particle excitations behave differently when the lattice site is already single-occupied or vacant, which can be characterized by two operators $d_{i\sigma} \equiv f_{i\sigma} n_{i\bar \sigma}$ and $e_{i\sigma} \equiv f_{i\sigma} (1-n_{i\bar \sigma})$.~\cite{Mancini04} While $f_{i\sigma}$ is the ``bonding'' combination of these two types of excitations, $e_{i\sigma} +d_{i\sigma}$, the ``anti-bonding'' counterpart, $e_{i\sigma} - d_{i\sigma}= f_{i\sigma}(1-2n_{i{\bar \sigma}})=f_{i\sigma}(-1)^{n_{i{\bar \sigma}}}$, is ${\bar f}_{i\sigma}$. It can be easily checked that ${\bar f}_{i\sigma}$ is a canonical fermion operator, satisfying the anticommutation rule $[{\bar f}_{i\sigma},{\bar f}_{j\sigma'}^\dag]_+=\delta_{ij}\delta_{\sigma\sigma'}$, and is orthogonal to $f_{i\sigma}$ in the single-occupancy limit, $\langle[{ f}_{i\sigma},{\bar f}_{i\sigma}^\dag]_+\rangle=\langle 1-2n_{i{\bar \sigma}}\rangle=0$. We examine the spectral function of this composite fermion operator in the two-impurity Anderson model,~\cite{ZhuSq11} which is solved {\it exactly} by the numerical renormalization group (NRG) method.~\cite{Wilson75,Krishnamurthy80,Bulla08} This model is not only relevant to the setting of double quantum dots connected to metallic leads, but also can be mapped from a lattice model, such as the Hubbard model or periodical Anderson lattice model, with a two-site cluster DMFT approach. It captures the ``Kondo-breakdown'' physics of lattice systems due to the competition between local quantum fluctuations such as Kondo dynamics, and short-range spatial fluctuations, such as the inter-site spin or change correlations.  With the advantages of the NRG method in its low-energy resolution and its dealing with real frequencies directly, we obtain also an accurate form for self-energies over the entire energy range for a model with both local and nonlocal interactions. We show that the self-energy for Anderson orbitals becomes singular (with a pole) near the Fermi energy in the regime dominated by the  inter-site spin exchange interaction [Ruderman-Kittel-Kasuya-Yosida (RKKY) interaction] where the Kondo fixed point is prevented. We further discover that instead of exhibiting unscreened local moments behavior, the Anderson orbitals develop a collective excitation mode in terms of the composite fermion ${\bar f}_{i\sigma}$. Specifically, we find that in the Kondo resonance regime dominated by onsite Kondo coupling, with a  Kondo resonance peak in $f_{i\sigma}$, the spectral functions of ${\bar f}_{i\sigma}$ have a pseudogap; in the regime dominated by RKKY interaction, where there is a pseudogap in excitations of $f_{i\sigma}$, the excitations of ${\bar f}_{i\sigma}$ have a ``resonance''-type spectrum.  We argue that ${\bar f}_{i\sigma}$, as a canonical fermion, could develop a Fermi surface but ``hidden'' from charge excitations.  The scattering off this composite fermion excitation mode leads to a pseudogap (or singularity in self-energy) of $f_{i\sigma}$. 

The rest of the paper is organized as follows. In Sec.~\ref{sec:model}, we introduce the two-impurity Anderson model, as an example system showing ``Kondo-breakdown'' effect due to the competition between onsite Kondo dynamics and intersite RKKY interaction, and describe the NRG method we adopt to solve this model. In Sec.~\ref{sec:singularity}, we show the results of the self-energy of Anderson orbitals in both the Kondo and RKKY-dominated regimes. In particular, we show that the self-energy develops a singularity (a pole structure) near the Fermi energy in the RKKY-dominated regime.  In Sec.~\ref{sec:cfermion}, we study the properties of the composite fermion operator ${\bar f}_{i\sigma}$ and show that the singularity is associated with a coherent excitation mode of the composite fermion. In Sec.~\ref{sec:explain}, we attempt to explain the physical mechanism for the composite fermion excitations. A summary is provided in Sec.~\ref{sec:summary}. More discussions are presented in the Appendixes, including the composite fermion excitations in particle-hole asymmetric cases, and the relation between composite fermion excitations and the Cooper pair excitations. 

\section{Model and Methods}
\label{sec:model}
 
We consider the two-impurity Anderson model, the Hamiltonian of which can be written as 
\begin{eqnarray}
H &=&  H_c + H_{cf}+H_f ,\nonumber \\
H_c&=& \sum_{{\bf k}\sigma} (\epsilon_{\bf k}-\mu) c^\dag_{{\bf k}\sigma} c_{{\bf k}\sigma}, \nonumber \\
H_{cf} &=&\frac{1}{\sqrt{N_{L}}}\sum_{{\bf k}\sigma i} (V_{\bf k}e^{i {\bf k}\cdot {\bf r}_i}
c^\dag_{{\bf k}\sigma}f_{i\sigma} + \text{H.c.}), \nonumber \\
H_f&=& \sum_{i\sigma} (\epsilon_f-\mu)  f^\dag_{i\sigma} f_{i\sigma} +\sum_i U n_{f i\uparrow}n_{f i\downarrow},
\label{eq:hamiltonian}
\end{eqnarray}
where $f_{i\sigma}$ ($i=1,2$) are two Anderson obitals with onsite  Coulomb repulsion $U$ and the energy level $\epsilon_f$. They hybridize at each site with non-interacting electrons $c_{\mathbf{k}\sigma}$ which form a band with dispersion $\epsilon_{\mathbf{k}}$. Here, $V_{\bf k}$ is the hybridization strength and $\mu$ is the chemical potential. $N_L$ is the lattice size of the conduction electron.  In cluster-DMFT approaches to the Anderson lattice model or one-band Hubbard model, $f_{i\sigma}$ are chosen from neighboring sites in bipartite sublattices with $c_{{\bf k}\sigma}$ describing an effective fluctuating bath. In terms of the even ($e$) or odd ($o$) parity combinations of the local orbitals, $f_{p=(e,o)\sigma} = (f_{1\sigma}\pm f_{2\sigma})/\sqrt{2}$, the bath spectral spectrum $\Delta_{e,o}(\omega)  =   \sum_{\bf k} (V^2_{\bf k}/2N_{L}) |e^{i{\bf k}\cdot {\bf r}_1} \pm e^{-i{\bf k}\cdot {\bf r}_2}|^2 /(\omega -\epsilon_{\bf k} + i 0^+)$, is to be determined by a self-consistent procedure. In this mapping, the even and odd parity states correspond to  the momentum points $\Gamma = (0,0,\ldots)$ and $M=(\pi,\pi,\ldots)$ for a $d$-dimension hypercubic lattice, respectively. 

In the local moment limit, $\epsilon_f \ll \mu$ and $\epsilon_f +U \gg \mu$, it is suggested that the zero and double occupancy configurations of Anderson orbitals can be projected out and the low-energy excitations can be described in terms of spin interactions~\cite{Jones87}
\begin{equation}
H_{cf}+H_{f} \to \sum_i J_K {\bf S}_i \cdot {\bf s}_i + I_0 {\bf S}_1 \cdot {\bf S}_2.  
\label{eq:H2K}
\end{equation} 
Here, the Kondo coupling $J_K\sim V^2/U$ couples the spins of the Anderson orbitals (${\bf S}_i$) and conduction electrons (${\bf s}_i$) at each impurity site.  The RKKY coupling $I_0$ is the intersite spin-exchange coupling between Anderson orbitals, which can be perturbatively generated $I_0 \sim J_K^2/D$,~\cite{ZhuSq11}  where $D$ is the bath electron bandwidth and is set as the energy unit ($D=1$).

We adopt the NRG method to solve the two-impurity Anderson model. The calculation details can be found in Ref.~\onlinecite{ZhuSq11}. In particular, we calculate various zero-temperature dynamical quantities,  $G_{ab} \equiv \langle \langle a; b^\dag \rangle\rangle$, where $a$ and $b$ are electron operators.  In NRG, the imaginary parts of dynamical quantities are directly calculated with the Lehmann representation (see Appendix ~\ref{sec:sign} for more details). The real parts can be determined subsequently from the Kramers-Kronig relation. We choose $a,b$ to be $d_{p\sigma}$ or $e_{p\sigma}$ to examine their individual behaviors. This not only enables us to determine the Green's functions of the regular and composite fermions: $G_{{ff}}$ ($G_{{\bar f}{\bar f}}$) = $(G_{ee}+G_{dd})\pm 2G_{de}$ (it is found that $G_{ed}=G_{de}$), but also provides a direct calculation for the self-energy of Anderson orbitals $\Sigma_{p\sigma}(\omega)$. The latter can be obtained from the equation of motion, 
\begin{eqnarray}
G^{-1}_{ff,p\sigma}(\omega) &= &\omega + \mu -\epsilon_f - \Delta(\omega) -\Sigma_{p\sigma}(\omega), \nonumber \\
\Sigma_{p\sigma}(\omega) &=&  \langle\langle [f_{p\sigma}, H_{f,int}] ; f^\dag_{p\sigma} \rangle\rangle /G_{ff,p\sigma}(\omega) \nonumber \\
&=&\frac U2 \left[ 1 - G_{{\bar f}f,p\sigma}(\omega) /G_{ff,p\sigma}(\omega) \right] \;, 
\label{eq:selfenergy}
\end{eqnarray}
where $ H_{f,int} = \sum_i U n_{f i\uparrow}n_{f i\downarrow}$,~\cite{footnote-se} and $G_{{\bar f}f}(\omega)= G_{ee}-G_{dd}$.  

\section{Singlularity in self-energies}
\label{sec:singularity}
 
We start from a presumed form of the bath spectrum, $\Gamma_{e,o}(\omega) = -\text{Im} \Delta(\omega) = \Gamma_0$ for $|\omega| \le D$. This case has been studied earlier in Refs.~\onlinecite{Sakai89} and \onlinecite{ZhuSq11}. We follow the same numerical procedure and adopt the same parameters $\Gamma_0 = 0.045\pi$, $U=-2\epsilon_f = 2$, and $\mu=0$ as in Ref.~\onlinecite{ZhuSq11} [as Case (i)]. As the generated RKKY interaction vanishes for this spectrum, we add an explicit intersite spin exchange term $I{\bf S}_1\cdot {\bf S}_2$ to simulate the RKKY interaction effect. This model exhibits a continuous transition from a Kondo resonance state to an inter-impurity singlet state with the tuning of the ratio of $I/T_K^0$.~\cite{Jones87,Sakai89,ZhuSq11}  
It is found that for $I<I_c\approx 2.3 T_K^0$, where $T_K^0\approx 10^{-3}$ as the single-ion Kondo temperature, the ground state is a Kondo resonance state governed by the Kondo strong-coupling fixed point, $J_{K,\text{eff}}\to \infty$. The spectral functions for Anderson orbitals have finite spectral weight $A_{f}(\omega)= -\text{Im} G_{ff}(\omega)=1/\Gamma_0$ at $\omega=0$ but with gradually reducing Fermi liquid (FL) temperature. For $I>I_c$, while the Kondo renormalization is cut off by the RKKY interaction,~\cite{ZhuSq11} the Kondo fixed point is prevented.  The spectral functions have vanishing spectral weight $A_{f}(\omega) \sim \omega^2$ near the Fermi energy. At the quantum critical point (QCP) $I=I_c$, various correlation functions are found to be divergent, including the staggered magnetic susceptibility, the inter-site singlet Cooper pair correlation function, and the current fluctuation between two sites.~\cite{Sakai89,ZhuSC}

\begin{figure}[tbh]
\centering
\includegraphics[width=\columnwidth]{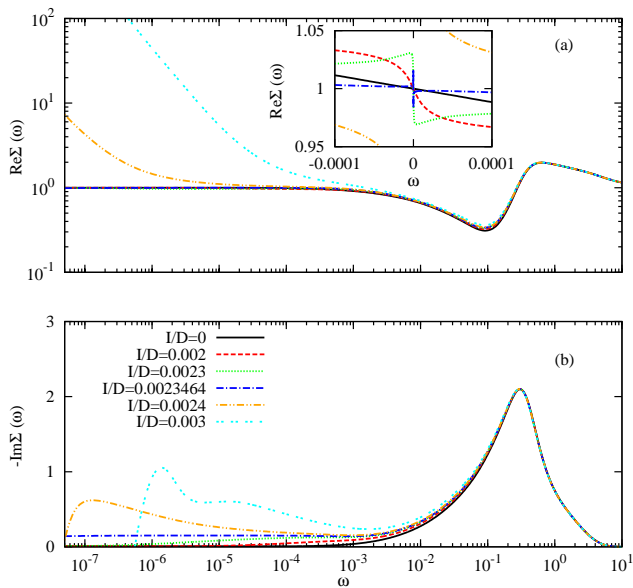}
\caption{(Color online) Real (a) and imaginary (b) parts of the self-energy for $f$ electrons for the case of $\Gamma_{e,o}(\omega) = \Gamma_0$ with added $I{\bf S}_1\cdot {\bf S}_2$ term. Here, $\Gamma_0 = 0.045\pi$, $U=-2\epsilon_f = 2$, and $\mu=0$. Various quantities for different parity and spin channels are the same due to the symmetry. Additional particle-hole symmetry is also preserved. In the inset of panel (a), we show the zoom-in region near the Fermi energy in linear-linear scale to demonstrate the behavior of  $\text{Re}\Sigma$ in the Kondo resonance regime. }
\label{fig:se1}
\end{figure}

In Fig.~\ref{fig:se1}, we show the results of self-energies for different values of $I$ [see also Fig. 3 for $A_{f}(\omega)$]. Here, the symmetry between even and odd channels is preserved, $G_{e\sigma} = G_{o\sigma}$. Therefore, in the site basis, $G_{(11),\sigma} = G_{(22),\sigma} = G_{e\sigma}$ while $G_{(12),\sigma} = 0$.  In the Kondo resonance regime ($I<I_c$), it is found that the self-energy has an analytical form $\text{Re}\Sigma_{p\sigma}(\omega) = U/2 - \omega/Z$ [see inset of Fig. 1(a)] and $\text{Im}\Sigma_{p\sigma}(\omega) \sim - \omega^2$ at low energies, while $Z \to 0$ approaching the QCP.  In the RKKY dominated regime ($I>I_c$), where a pseudogap form for the single-particle spectra is identified, we find $\text{Re}\Sigma_{p\sigma}(\omega) = Z'/\omega$ with $Z' \to 0$ approaching the QCP, i.e., the self-energy has a pole at the Fermi energy.~\cite{footnote} 
Apparently, the development of a pseudogap is associated with this singularity in the self-energy, as both $\text{Re}G_{ff,p\sigma}(\omega)$ and $\text{Im}G_{ff,p\sigma}(\omega) $ vanish.

\begin{figure}[tb]
\centering
\includegraphics[width=\columnwidth]{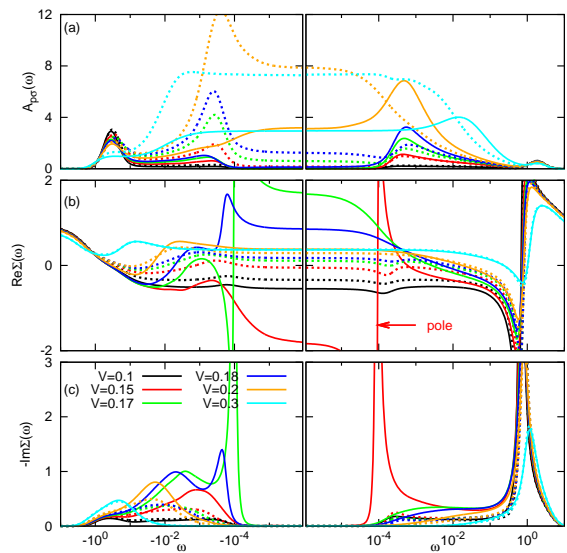}
\caption{(Color online) (a) Spectral function $A_f(\omega)$, (b) real and (c) imaginary parts of the self-energy for different values of the hybridization constant $V$ for a realistic band model. Here, $U=2$, $\epsilon_f =-0.5$, and $\mu =-0.2$. As $V$ increases, the relative ratio between the generated RKKY interaction $I_0$ and $T_K^0$ decreases, and the system evolves from the RKKY dominated regime to the Kondo resonance regime. The solid and dashed lines represent the even and odd parity channels, respectively. The spin symmetry is preserved. The pole position in $\text{Re}\Sigma$ for $V=0.15$ is illustrated as example. Correspondingly,  $\text{Im}\Sigma$ has a $\delta$-function-like peak. }
\label{fig:se2}
\end{figure}

In Fig.~\ref{fig:se2}, we further show the self-energies for the case with the bath spectrum determined from a realistic three-dimensional (3D) tight-binding dispersion for conduction electrons. Here, it is found that $\Gamma_{e,o}(\omega)$ have the form $\Gamma_0 (1\mp \omega)$ at low energies, and a finite antiferromagnetic RKKY interaction is perturbatively generated $I_0\approx 0.20 \rho_0 J_K^2$.~\cite{ZhuSq11} Therefore, we turn off the explicit spin exchange term $I{\bf S}_1\cdot {\bf S}_2$, and simply change the value of the hybridization constant $V=V_{\bf k}$ (with $U$ fixed) to tune the relative strength between $I_0$ and $T_K^0$.  The results in Fig. \ref{fig:se2} are for $U=2$, $\epsilon_f =-0.5$, and $\mu =-0.2$. Due to the lifting of the symmetry between even and odd parity channels, there is no sharp transition in this case. Instead, the system changes smoothly from a Kondo resonance state to an inter-impurity singlet state, with finite and almost vanishing single-particle spectral weights near the Fermi energy, respectively. We also identify the singularity in the self-energy of $f$ electrons in a certain range of the RKKY-dominated regime. In this case, the singularity (as a pole) only exists in the even parity channel and its position is shifted slightly away from the Fermi energy. 

\section{Composite fermion excitations}
\label{sec:cfermion}

\begin{figure}[tbh]
\centering
\includegraphics[width=\columnwidth]{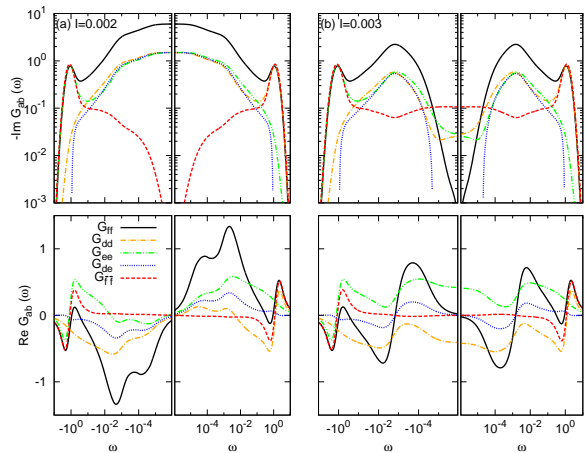}
\caption{(Color online) Imaginary and real parts of different composite fermion Green's functions, $G_{ff}$ (black), $G_{{\bar f}{\bar f}}$ (red), $G_{dd}$ (orange), $G_{ee}$ (green), and $G_{de}$ (blue). This is in correspondence to the case in Fig.~\ref{fig:se1} for two values of $I=0.002$ (a) and $I=0.003$ (b), as representatives of the Kondo resonance regime and the RKKY-dominated regime, respectively. The Green's functions for different parity ($p$) and spin ($\sigma$) channels are the same in this specific case.  $-\text{Im}\langle\langle d_{p\sigma}, e^\dag_{p\sigma}\rangle\rangle$ is negative in the energy range $|\omega|>U/2$, and additionally near $\omega=0$ for $I=0.003$ (not shown due to the log plot). 
} 
\label{fig:cf1}
\end{figure}
 
From Eq.~(\ref{eq:selfenergy}), the emergence of a pole in the self-energy takes place when $G_{ff,p\sigma}(\omega)$ vanishes while $G_{{\bar f}f,p\sigma}(\omega)$ is finite. A pole in self-energies implies the existence of another coherent excitation mode which $f_{p\sigma}$ is hybridized with. To seek this extra excitation mode, we examine the individual properties of composite fermion operators $d_{p\sigma}$  and $e_{p\sigma}$. 
In Fig.~\ref{fig:cf1}, we show the properties of these composite fermions as well as their hybridizations for the case in Fig.~\ref{fig:se1}.   In the Kondo resonance regime ($I=0.002$),  the excitations of composite fermions $d_{p\sigma}$ and $e_{p\sigma}$ are coherently hybridized in the form of the original fermion operator $f_{p\sigma}$. It is found that for each parity and spin channel, $\text{Im} G_{de}\approx \text{Im} G_{dd}\approx \text{Im}G_{ee}= \text{Im}G_{ff}/4$ at $\omega=0$. Together with both vanishing $\text{Re} G_{de}$ and  $ \text{Re} G_{dd}$, this implies that the hybridization between $d_{p\sigma}$ and $e_{p\sigma}$ is the coherent part of the single-electron excitation $f_{p\sigma}$ at the Fermi energy. This is consistent with the non-interacting nature of the strong-coupling fixed point, where $J_K\to \infty$ is equivalent to $U\to 0$.  Interestingly, their ``anti-bonding'' part ${\bar f}_{p\sigma} = e_{p\sigma}-d_{p\sigma}$ has a pseudogap form, $A_{\bar f}(\omega) =-\text{Im} G_{{\bar f}{\bar f}}(\omega)\sim \omega^2$. In the RKKY dominated regime ($I=0.003$), where $G_{ff}(\omega)$ has a pseudogap, $A_f \sim \omega^2$, we find that $A_{\bar f}(\omega)$ is finite near the Fermi energy. The latter resembles a ``Kondo resonance''. In this case, we find that $\text{Im} G_{de}$ has an opposite sign than $\text{Im} G_{dd}$, in contrast to the Kondo resonance regime. In addition, $\text{Re} G_{dd}(0)$ and $\text{Re} G_{ee}(0)$ take finite values of the order of $ \mp 1/U$. 

Specifically, the relative sign between the matrix elements $\langle n|e^\dag_{e\sigma}|G\rangle$ and $\langle n|d^\dag_{e\sigma}|G\rangle$ between the ground state ($G$) and an excited state ($n$) is different in two regimes: they have the same sign in the Kondo resonance regime, but have an opposite sign in the RKKY dominated regime. As 
\begin{equation}
-\text{Im}G_{ff} (G_{{\bar f} {\bar f}})\sim |  \langle n|e^\dag_{e\sigma}|G\rangle \pm \langle n|d^\dag_{e\sigma}|G\rangle|^2,
\label{eq:lehm}
\end{equation}
the constructive/destructive interference behavior between $e_{p\sigma}$ and $d_{p\sigma}$ excitations leads to the resonance-pseudogap ``dual'' relationship in two regimes.  We show further details and results in Appendix ~\ref{sec:sign}. 

Similar features can also be identified in a realistic band model, as shown in Fig.~\ref{fig:se2}. However, in cases where the particle-hole symmetry is broken (away from single-occupancy), we find that the resonance-pseudogap ``duality'' between $f_{p\sigma}$ and ${\bar f}_{p\sigma}$ is captured instead by the new operators $(g_{p\sigma}, {\bar g}_{p\sigma} )= \alpha e_{p\sigma} \pm \beta d_{p\sigma}$ with $\alpha/\beta\neq 1$ determined from the occupancy number (see Appendix ~\ref{sec:phasym}).  Here, $f_{p\sigma}$ and ${\bar f}_{p\sigma}$, as projections from $g_{p\sigma}$ and ${\bar g}_{p\sigma}$, could both have finite spectral weight. But, the sign change in $\text{Im} G_{de}$ is a robust feature in the transition (or crossover) point (see Appendix ~\ref{sec:phasym}). 

\section{Discussions}
\label{sec:explain}

The above results show that the excitations of $d_{p\sigma}$ and $e_{p\sigma}$ are no longer equivalent in presence of Coulomb interaction. In other words, if we try to formulate a quasiparticle model from $f_{p\sigma}$, it is necessary to incorporate its orthogonal mode ${\bar f}_{p\sigma}$, as $\langle [{\bar f}_{p\sigma}, f^\dag_{p\sigma}]_+\rangle = 0$ in the single-occupancy limit.  The form of $\Sigma_{p\sigma}$ [cf. Eq.~(\ref{eq:selfenergy})] also implies that an effective hybridization with ${\bar f}_{p\sigma}$ provides a {\it full} account of the interaction effects. Indeed, we find the above results can be schematically explained by an effective hybridization model: 
\begin{equation}
{\hat G}^{-1} = 
\left(
\begin{array}{cc}
G_0^{-1} & -U/2 \\
-U/2 & {\tilde G}_0^{-1}
\end{array}
\right)
\label{eq:hybmodel}
\end{equation}
in the basis ($f_{p\sigma}$,${\bar f}_{p\sigma}$), where $G_0^{-1} = \omega+\mu -\epsilon_f-U/2 - \Delta(\omega)\approx \omega + i\Gamma_0$.  However, due to the composite-operator nature of ${\bar f}$, it is not straightforward to determine ${\tilde G}_0^{-1}$. We can learn instead its behavior from $\Sigma_{ff}= (U/2)^2 {\tilde G}_0$ [i.e., $\Sigma_{p\sigma}-U/2$ defined in Eq.~(\ref{eq:selfenergy})]. In the Kondo resonance regime,  ${\tilde G}_0 \sim \Sigma_{ff}\sim a\omega - i b \omega^2$. This is in a form of incoherent fermion excitations devoid of any poles near the Fermi energy.  ``Hybridizations'' with these excitations provide Fermi-liquid-type corrections to the the regular fermion excitations $f_{p\sigma}$.    
In the RKKY dominated regime,   ${\tilde G}_0 \sim 1/(c\omega +i0^+)$, which is a form of a coherent free particle. ``Hybridization'' with such a mode produces the pseudogap spectra for $G_{ff}$. From $G_{{\bar f}{\bar f}}=1/[{\tilde G}_0^{-1} - (U/2)^2/G_0^{-1}]$, we can obtain $-\text{Im}G_{{\bar f}{\bar f}}(0) =1/\Gamma_0'= 4\Gamma_0/U^2\approx 0.14$, consistent with the numerical results. The relation $\Gamma_0 \Gamma_0' = (U/2)^2$ has been further verified by other values of $V$ and $U$. 

The emergence of the free-particle form of the composite fermion ${\tilde G}_0 \sim 1/(\omega +i0^+)$, as well as the property $\text{Re}G_{ee}(0), \text{Re}G_{dd}(0) \sim \pm 1/U$ in the RKKY dominated regime,  are reminiscent of those in the atomic limit, where, for $\epsilon_f = -U/2$,  
\begin{eqnarray} 
H_f &=& \sum_{i\sigma}\frac U2 (- e^\dag_{i\sigma}e_{i\sigma} +  d^\dag_{i\sigma}d_{i\sigma})=
 \sum_{i\sigma}\frac U2 {\bar f}_{i\sigma}^\dag f_{i\sigma},  
\label{eq:hfenergy}
\\
G_{ff,ii} &=&  \frac{ 1/2}{ {\omega +U/2}} + \frac{ 1/2 }{ {\omega-U/2}}  = \frac{1 }{ \omega -{(U/2)^2 / \omega}}. 
\end{eqnarray}
The two poles in the Green's function $\mp U/2$ correspond to the ``fractionalized'' excitations of $e$ and $d$ operators, respectively. This could also be effectively treated as a ``hybridization'' effect between $f$ and ${\bar f}$, where $G_0={\tilde G}_0 = 1/(\omega +i0^+)$. We argue in the following that, the coupling between $f$ electrons with a conduction electron band can lead to a finite spectral weight for ${\bar f}$ while $f$ keeps a pseudogap form if the Kondo fixed point is prevented. 

NRG studies show that the RKKY dominated regime and the Kondo resonance regime have similar low-energy properties.~\cite{Jones87,Sakai89,ZhuSq11} Both of them belong to (local) Fermi-liquid fixed points, and the low-energy excitations such as spin and charge susceptibilities of Anderson orbitals show Fermi-liquid behaviors.~\cite{Sakai89,ZhuSq11}  This can also be evidenced from the same structure of the low-energy effective Hamiltonian, from which Landau parameters are deduced.~\cite{Jones87} They differ by a scattering phase shift, $0$ or $\pi/2$, or effective electron numbers (modes) in the ground state, which leads to distinct single-particle excitations of Anderson orbitals.  Specifically,  NRG formulates the non-interacting conduction electron band as a semi-infinite chain of electron modes with nearest-neighbor hopping, ~\cite{Wilson75,Krishnamurthy80,Bulla08} 
\begin{equation}
H_c =  \sum_{n=0}^{\infty}\sum_{\sigma} (t_n f^\dag_{n\sigma} f_{n+1,\sigma}+\text{H.c.}) .
\label{eq:nrgchain}
\end{equation}
The hybridization between an Anderson orbital $f_{\sigma}$ and the conduction electron band can be treated as adding a site to the chain,
\begin{equation}
H_{cf} = {V} \sum_{\sigma} (f^\dag_{-1,\sigma} f_{0\sigma} +\text{H.c.}), 
\label{eq:hcf} 
\end{equation}
where $f_{-1,\sigma} \equiv f_{\sigma}$, and $f_{0\sigma}$ is the conduction electron operator at the impurity site. The Coulomb interaction term, as the ``hybridization'' between $f$ and ${\bar f}$, from  Eq. (\ref{eq:hfenergy}),  can be treated as adding another site to the chain, 
\begin{equation}
H_f =  (U/4) \sum_{\sigma} (f^\dag_{-2,\sigma} f_{-1,\sigma} +\text{H.c.}),
\label{eq:hf}
\end{equation}
where  $f_{-2,\sigma} \equiv {\bar f}_{\sigma}$.  If $H_c$ alone is diagonalized by the NRG iterative procedure, it is found that the spectra at even and odd iterations are distinct, corresponding to two FL fixed points with odd and even number of electrons (modes) in the ground state.~\cite{Krishnamurthy80} When Anderson orbitals are added, it is found that in the Kondo resonance regime, the NRG spectra at a large even/odd iteration are the same as those at an odd/even iteration for $H_c$. This can be understood as that, while the Kondo strong-coupling fixed point is reached, $J_{K,\text{eff}}\to \infty$ or $U_{\text{eff}}\to 0$, $f_{-2,\sigma}$ is decoupled and $f_{-1,\sigma}$ is added to the chain as a non-interaction electron mode. Here, we obtain a resonance form for $f_{-1}$ as the Kondo resonance, but a pseudogap form for $f_{0}$. The resonance weight at the Fermi energy takes the form $1/\Gamma_0\sim (D/V)^2/D$, where $D$ and $V$ are the hopping matrix elements for $f^\dag_0 f_1$ and $f^\dag_{-1}f_0$, respectively. In the RKKY dominated regime, however, the spectra at a large even/odd iteration remain the same as those at an even/odd iteration for $H_c$. As the Kondo strong-coupling fixed point is prevented, or $U_{\text{eff}}$ remains finite, $f_{-2}$ is also effectively coupled to the chain, contributing an extra mode to the chain. In analogy to the Kondo resonance regime, we expect $f_{-2}$, now as the head site, to develop a resonance form in spectral function while the excitations of $f_{-1}$ have a pseudogap. The resonance weight for $f_{-2}$ would follow $1/\Gamma_0'\sim (V/U)^2/D$, where $V$ and $U$ are the hopping matrix elements for $f^\dag_{-1} f_{0}$ and $f^\dag_{-2}f_{-1}$, respectively. This is indeed in agreement with the numerical results on ${\bar f}$. Here, $f_{-2}$, taking the place of $f_{-1}$ as the low-energy ``quasiparticle'' excitations, contributes to various FL properties. 

We further notice that the chain modes $f_{n\sigma}$ in Eq.~(\ref{eq:nrgchain}) are constructed in alternative ``bonding'' and ``anti-bonding'' combinations of negative and positive energy modes, similar to the definition  $f_{-1\sigma}(f_{-2\sigma}) = e_{\sigma}\pm d_{\sigma}$ [cf. Eq.(\ref{eq:hfenergy})]: 
\begin{eqnarray}
f_{n\sigma} &=& \sum_{m=0}^{\infty} c_\Lambda(m,n) (a_{m\sigma} - b_{m\sigma}), \text{ if $n$ is even}; 
\nonumber \\
  &=&\sum_{m=0}^{\infty} c'_\Lambda(m,n) (a_{m\sigma} + b_{m\sigma}), \text{ if $n$ is odd},
\label{eq:fevenodd}    
\end{eqnarray}
where $a_{m\sigma}$ and $b_{m\sigma}$ are the conduction electron modes ($s$-wave part) in the negative and positive energy grids [$\mp D\Lambda^{-(m+1)}$,  $\mp D\Lambda^{-m}$] ($\Lambda>1$ is a discretization parameter), and  
\begin{eqnarray}
H_c &=& \sum_{\sigma} \int_{-D}^{D} d \epsilon \epsilon c^\dag_{\epsilon\sigma}c_{\epsilon\sigma}, \nonumber \\
&\to& \sum_{m=0}^{\infty} \sum_{\sigma} \epsilon_m (-a^\dag_{m\sigma}a_{m\sigma} + b^\dag_{m\sigma}b_{m\sigma}). 
\label{eq:gridmodes}
\end{eqnarray}
If we treat the energy $\epsilon$ in Eq.~(\ref{eq:gridmodes}) as momentum $k-k_F$ in 1-D, we realize that different combinations $a_{m\sigma}\pm b_{m\sigma}$ are related to different parity combinations of two chiral modes  $\psi_{L,R}(x) \sim \int e^{\pm i (k-k_F)x}c_k dk$.~\cite{Affleck91} This may imply that the composite fermion operator ${\bar f}_{i\sigma}$ corresponds to a different parity combination of certain chiral degrees of freedom than regular fermion operators. These two types of excitations only couple if the effective interaction is finite, or the Kondo strong-coupling fixed point is prevented.   

We can also gain intuition on the physical meaning of the composite fermion operator from the sign structure in spin and charge (as a pseudospin) spaces. The spin operator for Anderson orbitals, defined from either the regular fermion operator or the composite operator, are the same: ${\bf S}_{i} = {\bar f}^\dag_{i\alpha} ({\vec \tau}_{\alpha\alpha'}/2){\bar f}_{i\alpha'}= {f}^\dag_{i\alpha} ({\vec \tau}_{\alpha\alpha'}/2){f}_{i\alpha'}$. However, for the pseudospin operators, which are defined as $J_i^+ = (-1)^i f^\dag_{i\uparrow}f^\dag_{i\downarrow}$, $J_i^-= (J_i^+)^\dag$, and $J_i^z = (\sum_{\sigma}f^\dag_{i\sigma}f_{i\sigma}-1)/2$, we find that $J_i^{\pm} = -{\bar J}_i^{\pm}$, where ${\bar J}_i^+\equiv (-1)^i {\bar f}^\dag_{i\uparrow} {\bar f}^\dag_{i\downarrow}$. (This also implies that in a slave-fermion representation to spins in spin Hamiltonians, it is not known {\it a priori} which fermion the slave fermion corresponds to, the regular fermion or the composite fermion. This should be decided by the pseudospin configurations.) 

For non-interacting electrons, the ground state of two electrons coupled by a hybridization or a hopping term, such as $H_{cf}=V\sum_{\sigma} (f^\dag_{-1\sigma}f_{0\sigma}+\text{H.c.})$, is  
\begin{equation}
\psi_G =  (\left|\uparrow_{-1}, \downarrow_0\rangle\right.-\left|\downarrow_{-1}, \uparrow_0\rangle\right.) +  (\left|\uparrow\downarrow_{-1}, 0_0\rangle\right.+\left|0_{-1}, \uparrow\downarrow_0\rangle\right.)
\label{eq:gsfree}
\end{equation}
at half-filling, which is a combination of a spin-singlet state and a pseudospin-singlet state. We notice that in the Schrieffer-Wolff transformation to map the Anderson-type Hamiltonians into Kondo-type Hamiltonians, in addition to the spin Kondo coupling term $J_K{\bf S}_{-1} \cdot {\bf S}_0$, there is also a charge Kondo coupling term
\begin{eqnarray}
H_{ck} &=& -\frac {J_K }{4} n_{f_{0}} n_{f_{-1}} + \frac{J_K }{ 2} (f^\dag_{-1\uparrow}f^\dag_{-1\downarrow} f_{0\downarrow} f_{0\uparrow} + h.c) \nonumber \\
&=& -J_K {\bf J}_{-1} \cdot {\bf J}_0.
\label{eq:chargekondo}
\end{eqnarray} 
For repulsive interaction $U>0$, the charge Kondo coupling is ``ferromagnetic'' and favors pseudospin-triplet configuration.  The Kondo renormalization in some way can be understood as to ``overcome'' the mismatch between spin and pseudospin spaces. (Similarly, for attractive interaction $U<0$, the roles of spin and pseudospin couplings exchange and the system renormalizes into a strong coupling charge Kondo fixed point.) This term is usually neglected when the zero and double occupancy states have vanishing weight and are projected out in the $U\to \infty$ limit. However, when calculating directly the spectral functions for Anderson orbitals, we need to keep the  zero and double occupancy configurations, and therefore, to consider all couplings at finite $U$. When the Kondo strong-coupling fixed point is prevented, or $J_K$ remains finite, we expect that the charge Kondo coupling term, as a marginal irrelevant parameter, affects the single-particle excitations. In terms of antiferromagnetic spin coupling and ``ferromagnetic'' pseudospin coupling,  a ground state with combinations of spin-singlet and pseudospin triplet is favored. On the other hand, as ${\bar J}_{-1}^+ = -J_{-1}^+$, the pseudospin triplet between $f_{-1}$ and $f_{0}$ becomes a singlet between ${\bar f}_{-1}$ and $f_{0}$. Therefore, this ground state is rather described by an effective coupling $H=\sum_{\sigma} ({\bar f}^\dag_{-1\sigma}f_{0\sigma}+h.c)$, and a resonance form in composite fermion operator follows.  This also implies that the effective Hamiltonian in the low-energy region of the RKKY dominated regime is similar to that in the local moment fixed point. For example, we observe the spectral weight of the composite fermion at Fermi energy in the RKKY dominated regime is about the same as its weight at $|\omega|\approx 0.3$ (cf. Fig.~\ref{fig:cf1}), which is controlled by the local-moment fixed point. 

\section{Conclusions}
\label{sec:summary}

In summary, we have studied the single-particle excitation properties of the two-impurity Anderson model, as an example of interacting electron systems with competing local and non-local interactions. We show that in the RKKY dominated regime, where the Kondo strong-coupling fixed point is prevented (or Kondo breakdown), the excitations of regular fermion operators $f_{i\sigma}$ for Anderson orbitals show a pseudogap form. Correspondingly, the self-energy is singular. As the pole in self-energies commonly indicates another collective mode the regular fermion operator couples to, we trace down the mode as excitations of a composite fermion operator ${\bar f}_{i\sigma} = f_{i\sigma} (-1)^{n_{i{\bar \sigma}}}$.  We show that its spectral function develops a ``resonance'' form in the RKKY dominated regime. In the Kondo resonance regime, where the regular fermion operators develop a Kondo resonance, the spectral function for composite fermion operators has a pseudogap form. 

Although the physical meaning of the composite fermion operator is currently not very clear to us, we attempted to provide some discussions on its nature and the resonance-pseudogap ``dual'' -relationship between the regular fermion operators and the composite fermion operators. In interacting systems, as evidenced by our calculations, the single-particle excitations behave differently when a lattice site is vacant or already single occupied, which can be characterized by the electron operators $e^\dag_{i\sigma}$ and $d^\dag_{i\sigma}$. To provide a full account of the single-particle excitations $f^\dag_{i\sigma} = e^\dag_{i\sigma}+d^\dag_{i\sigma}$ , it is necessary to introduce the composite fermion operator  ${\bar f}^\dag_{i\sigma} = e^\dag_{i\sigma}-d^\dag_{i\sigma}$ to account for their difference.  We notice that the ``bonding'' and ``anti-bonding'' relationship is analogous to the different combination between negative and positive energy modes $a_{m\sigma}\pm b_{m\sigma}$ of conduction electrons [cf. Eq.~(\ref{eq:fevenodd})]. The latter is associated with different parity combinations of ``left-moving'' and ``right-moving''  chiral degrees of freedom. While the resonance form of $f_{i\sigma}$ captures the fermionic excitations in the Kondo resonance regime with $\pi/2$ phase shift, the ``resonance'' form of ${\bar f}_{i\sigma}$ instead captures the fermionic excitations in $0$ phase shift limit. We show that the effective Hamiltonian in the RKKY dominated regime, for repulsive interaction, contains not only an anti-ferromagnetic spin Kondo coupling term, but also  a ``ferromangetic'' pseudospin (charge) Kondo coupling term [cf. Eq.~(\ref{eq:chargekondo})].  Such a combination is unfavorable for the regular fermion hybridization (hopping) $(f^\dag_{i\sigma} c_{i\sigma} +h.c.)$. However, the pseudospin operator in terms of the composite fermion ${\bar J}^\pm_{i\sigma}$ has an opposite sign as $J^\pm_{i\sigma}$ in terms of regular fermions, i.e., the pseudospin triplet state between $f_{i\sigma}$ and $c_{i\sigma}$ is a pseudospin singlet state between ${\bar f}_{i\sigma}$ and $c_{i\sigma}$. An effective hybridization (hopping) $({\bar f}^\dag_{i\sigma} c_{i\sigma} +h.c.)$ is favored. 

We find that the emergence of a coherent mode in composite fermions is a manifestation of the ``Kondo-breakdown'' effect, or Mott physics. The composite fermion, as an additional excitation mode, couples to regular fermion excitations and leads to a gapped behavior for the latter.  This takes place when the Kondo fixed point is prevented, or the effective interaction $U$ (as the coupling term) remains finite. In the two-impurity Anderson problem, this is due to the intersite spin-exchange coupling, as a ``cutoff'' on the Kondo renormalization.~\cite{ZhuSq11} Compared with the traditional understandings of this problem, the introduction of the composite fermions may bring the following technical and conceptual advantages. The composite fermion is a canonical fermion and by definition, it is a directly calculable quantity, at least in numerics.  Coherent modes of composite fermions provide a direct characterization of the ``Kondo-breakdown'' state.  It shows that not {\it all} fermionic excitations are gapped and naturally explains the Fermi-liquid behaviors of the spin and charge susceptibilities in the RKKY dominated regime $-\text{Im}\chi(\omega)\sim \omega$. In addition, the composite fermions may help to address the non-Fermi-liquid behaviors as well as finite Cooper pair fluctuations in the quantum critical regime. In Appendix~\ref{sec:pair}, we show that both the regular fermions and composite fermions have finite spectral weights in the quantum critical regime, but their weights are only the half of those in the Kondo resonance regime or the RKKY dominated regime, correspondingly.  We also show that the spectral weight for pair excitations (in an intersite spin-singlet configuration) is finite if and only if the spectral weights for both the regular fermions and composite fermions are finite.  

Our calculations have established that the composite fermions have coherent excitations in the ``Kondo-breakdown'' state of the two-impurity Anderson model. An important question arises  as to whether the concept of composite fermions can be generalized to the Mott-insulating states in lattice systems. We notice that in the Hubbard model or the Anderson lattice model, the interaction effects due to the onsite Coulomb interaction can be {\it fully} incorporated in ``hybridization'' between regular fermions and composite fermions, $H_U =\sum_{i} U(n_{i\uparrow}-1/2)(n_{i\downarrow}-1/2) = \sum_{i\sigma} (U/2) f^\dag_{i\sigma} {\bar f}_{i\sigma}=\sum_{{\bf k}\sigma} (U/2) f^\dag_{{\bf k}\sigma}{\bar f}_{{\bf k}\sigma}$, where ${\bar f}_{{\bf k}\sigma} = \sum_i {\bar f}_{i\sigma}e^{-i {\bf k}\cdot {\bf r}_i}= \sum_{\bf q} f_{{\bf k+q}\sigma} (\delta_{{\bf q},0}-2{n_{{\bf q}\bar \sigma}}) $, and $n_{{\bf q}{\bar \sigma}} = \sum_i e^{-i {\bf q}\cdot {{\bf r}_i}} n_{i{\bar \sigma}}$. If the composite fermion operators ${\bar f}_{{\bf k}\sigma}$ could have coherent excitations, or develop a ``Fermi surface'' at $\eta_{\bf k}$, this corresponds to a Luttinger surface where the self-energies are singular and the Green's function for regular fermions becomes vanishing, or $G({\bf k}, \omega) = 1/(\omega-\epsilon_{\bf k} -\frac{ {U}_{\text{eff}}^{2}}{\omega-\eta_{\bf k}}) = a_{\bf k}/(\omega- \epsilon_{{\bf k}1} )+ b_{\bf k}/(\omega-\epsilon_{{\bf k}2})$. Indeed, such a Luttinger surface has been identified in cluster DMFT calculations to the Hubbard model.~\cite{Tremblay04, Sakai10} It will be interesting to calculate the spectral functions of the composite fermion operator directly in these calculations to make the connection. We notice that this form of the single-particle Green's function has been proposed as a phenomenological theory for the pseudogap state in cuprate superconductors.~\cite{Yang06} We also notice that, in other proposals for the pseudogap state~\cite{Anderson,Yamaji11,Prelovsek02}, the basis is to introduce extra or ``hidden'' electronic excitation modes for interacting electrons.  

Although the definition of the composite fermions is theoretically straightforward, it remains a puzzle whether these excitations are physical. Even if there are coherent excitations or a ``Fermi surface'' of composite fermions, we expect that these excitations are not detectable or ``hidden'' from conventional spectra measurements. The reason is that the composite fermion operator is orthogonal to regular fermion operators. For example, in the tunneling spectra involving external source of electron $c^\dag_{\sigma}$,   $\langle c^\dag_{\sigma} {f}_{i\sigma}\rangle=\langle c^\dag_{\sigma} {\bar f}_{i\sigma}(-1)^{n_{i{\bar \sigma}}}\rangle$. Although ${\bar f}_{i\sigma}$ has finite excitations, the average vanishes as $\langle (-1)^{n_{i{\bar \sigma}}}\rangle= \langle 1-2n_{i{\bar \sigma}}\rangle\approx 0$ for  $\langle n_{i{\bar \sigma}}\rangle \approx 1/2$. A similar orthogonality issue is also raised in Ref.~\onlinecite{Anderson}. As $(-1)^{n_{i{\bar \sigma}}}$ acts as a $Z_2$ Ising spin, the charge carried by $f_{i\sigma}$ from $U(1)$ gauge transformation must be carried by ${\bar f}_{i\sigma}$ as well. As the system is commonly an insulator, it implies that the current carried by ${\bar f}_{i\sigma}$ responds to external electromagnetic fields in an anomalous way, resembling the  ``chiral anomaly''.  The solution may rely on the understandings of the gauge structure associated with the $Z_2$ Ising spin $(-1)^{n_{i{\bar \sigma}}}$, which captures the matching of Marshall signs~\cite{marshall55,lieb89,weng11} in spin and pseudospin configurations. We will leave it to a future study.  

\begin{acknowledgments} 
We thank A. V. Balatsky, T.-K. Ng, Q. Si, C. M. Varma and Z.-Y. Weng for helpful discussions. We also thank D. MacLaughlin for a critical reading of an early version of the manuscript. This work was supported by the NNSA of the U.S. DOE at LANL under Contract No. DE-AC52-06NA25396 (L.Z. \& J.-X.Z.), the LANL LDRD Program (L.Z. \& J.-X.Z.), and NSF grant DMR-0906530 (L.Z.). Part of the calculations were performed on a computer cluster at Center for Integrated Nanotechnologies, a U.S. DOE Office of Basic Energy Sciences user facility (Project No. C2011A1070).
\end{acknowledgments}

\appendix

\section{The matrix element of electron operators}
\label{sec:sign}

The calculation of the spectral functions follows the Lehmann representation 
\begin{equation} 
-\text{Im} G_{AB} (\omega\ge0) = \pi \sum_{n} \langle n |A^\dag|G\rangle \langle n |B^\dag|G\rangle \delta(\omega-E_n),  
\label{eq:lehmann}
\end{equation}
where $G$ is the ground state, and $n$ sums over all excited states with the energy $E_n$. In NRG, the Hamiltonian is diagonalized iteratively to incorporate gradually the low-energy sites. When the eigenspace becomes large, only a certain number of low-energy states are kept for later iterations. In traditional spectral function calculations, 
the excited states are chosen from kept states in even (or odd) iterations. However, states from different iterations are not necessarily orthogonal to each other and a patching scheme is commonly adopted.~\cite{Bulla08}  It is noticed that all the discarded states from all iterations are orthogonal and form a complete Fock space (CFS) conserving the full density matrix.~\cite{Anders06} We follow the CFS scheme in our calculations. The $\delta$-function in NRG is broadened in the log-Gaussian form,~\cite{Bulla08}
\begin{equation}
\delta(\omega-E_n) \to\frac {e^{-b^2/4} }{ b E_n \sqrt{\pi}} e^{-\ln^2(\omega/E_n)/b^2},
\end{equation}
where $b$ is a broadening parameter. For $\Lambda=2$, we take $b=0.6$, as commonly adopted.~\cite{Bulla08}

In our calculations, we calculate separately the contributions to spectral functions from $e^\dag_{i\sigma}$ and $d^\dag_{i\sigma}$, or single-particle excitations when the site is empty or single occupied. In addition, we examine each matrix element of $e^\dag_{i\sigma}$ and $d^\dag_{i\sigma}$ between different configurations of the Anderson orbitals. For two orbitals, there are 16 states, such as $\left|\uparrow_1,0_2\rangle\right.$. We further sort them according to their quantum numbers in charge, spin, and parity. An eigenstate can be written in the form 
\begin{equation}
| \Psi \rangle = \sum_{\alpha} a_{\alpha} |\alpha\rangle |\psi_{c\alpha}\rangle,
\end{equation} 
where $|\alpha\rangle$ is one of the 16 configurations of two Anderson orbitals, and $|\psi_{c\alpha}\rangle$ is the associated conduction electron configuration. The coefficient $a_{\alpha}$ gives the weight for each configuration and satisfies $\sum_{\alpha} a_{\alpha}^2=1$. Therefore, the matrix element of an electron operator can be written as 
\begin{eqnarray}
\langle n| A^\dag |G\rangle &=& \sum_{\alpha, \alpha'} M_{\alpha',\alpha}(A), \nonumber \\ 
M_{\alpha',\alpha}(A) &=& a^n_{\alpha'}a^G_{\alpha} \langle\psi^n_{c\alpha'}|\psi^G_{c\alpha}\rangle \langle \alpha' |A^\dag|\alpha\rangle. 
\end{eqnarray} 

We list the results for $M_{\alpha',\alpha}(A)$ between the ground state and a typical low-energy NRG excitation state in Table~\ref{tab:matrixelement}, for $I=0.002$ (Kondo resonance regime) and $I=0.003$ (RKKY dominated regime).  Here, the electron operators in the even channel and spin-up case are taken as an example, with  $e^\dag_{e\uparrow}\equiv {1\over\sqrt 2} (e^\dag_{1\uparrow}+e^\dag_{2\uparrow})$ and $d^\dag_{e\uparrow}\equiv {1\over\sqrt 2} (d^\dag_{1\uparrow}+d^\dag_{2\uparrow})$.  

We find that the difference between the Kondo resonance regime and the RKKY dominated regime is the sign change in $-\text{Im}G_{ed}$. Specifically, $ \langle n |e^\dag_{e\uparrow}|G\rangle$ and $ \langle n |d^\dag_{e\uparrow}|G\rangle$ have the same sign in the Kondo resonance regime while different signs in the RKKY dominated regime. This constructive/destructive behavior leads to resonance-pseudogap ``dual'' relations to $G_{ff}$ and $G_{{\bar f}{\bar f}}$.  

\begin{table*}[tbh]
\begin{tabular}{cc|cccc|ccccc}
\hline\hline 
&&\multicolumn{4}{c|}{I=0.002  ($E_n$ = 8.29E-8)} &  \multicolumn{4}{c}{I=0.003 ($E_n$ = 8.20E-8)}  \\
\hline
$\alpha$ & $\alpha'$ &   $M(e_{e\uparrow})$ & $M(d_{e\uparrow})$ & $M(f_{e\uparrow})$ & $M({\bar f}_{e\uparrow})$ &
  $M(e_{e\uparrow})$ & $M(d_{e\uparrow})$ & $M(f_{e\uparrow})$ & $M({\bar f}_{e\uparrow})$ \\ 
\hline 
$|0,0\rangle$ & ${1\over \sqrt{2}} (|\uparrow,0\rangle+|0,\uparrow\rangle)$ &
9.05E-6 & 0 & 9.05E-6 & 9.05E-6 &
-6.96E-7 & 0 & -6.96E-7 & -6.96E-7  \\
${1\over \sqrt{2}} (|\downarrow,0\rangle+|0,\downarrow\rangle)$ & 
${1\over \sqrt{2}} (|\uparrow,\downarrow\rangle-|\downarrow,\uparrow\rangle)$&
7.67E-5 & 0 & 7.67E-5 & 7.67E-5  &
-6.27E-5 & 0 & -6.27E-5 & -6.27E-5  \\
${1\over \sqrt{2}} (|\downarrow,0\rangle+|0,\downarrow\rangle)$ & 
${1\over \sqrt{2}} (|\uparrow\downarrow,0\rangle+|0,\uparrow\downarrow\rangle)$&
0 & 4.52E-6 & 4.52E-6  & -4.52E-6  &
0 & 3.52E-7 & 3.52E-7  & -3.52E-7  \\
${1\over \sqrt{2}} (|\uparrow,0\rangle-|0,\uparrow\rangle)$ & 
$|\uparrow,\uparrow \rangle$&
1.36E-4 & 0 & 1.36E-4  & 1.36E-4  &
2.68E-5 & 0 & 2.68E-5  & 2.68E-5  \\
${1\over \sqrt{2}} (|\downarrow,0\rangle-|0,\downarrow\rangle)$ & 
${1\over \sqrt{2}} (|\uparrow,\downarrow\rangle+|\downarrow,\uparrow\rangle)$&
6.80E-5 & 0 &  6.80E-5 & 6.80E-5  &
1.34E-5 & 0 &  1.34E-5 & 1.34E-5  \\
${1\over \sqrt{2}} (|\downarrow,0\rangle-|0,\downarrow\rangle)$ & 
${1\over \sqrt{2}} (|\uparrow\downarrow,0\rangle+|0,\uparrow\downarrow\rangle)$&
0 & 4.52E-6 &  4.52E-6 & -4.52E-6 &
0 & 3.52E-7 &  3.52E-7 & -3.52E-7 \\
${1\over \sqrt{2}} (|\uparrow\downarrow,0\rangle+|0,\uparrow\downarrow\rangle)$&
${1\over \sqrt{2}} (|\uparrow\downarrow,\uparrow\rangle+|\uparrow,\uparrow\downarrow\rangle)$ & 
4.52E-6 & 0 & 4.52E-6 &  4.52E-6  &
-3.48E-7 & 0 & -3.48E-7 &  -3.48E-7  \\
${1\over \sqrt{2}} (|\uparrow,\downarrow\rangle-|\downarrow,\uparrow\rangle)$&
${1\over \sqrt{2}} (|\uparrow\downarrow,\uparrow\rangle+|\uparrow,\uparrow\downarrow\rangle)$ & 
0 & 7.66E-5 & 7.66E-5 &  -7.66E-5  &
0 & 6.28E-5 & 6.28E-5 &  -6.28E-5 \\
$|\downarrow,\downarrow\rangle$&
${1\over \sqrt{2}} (|\uparrow\downarrow,\downarrow\rangle+|\downarrow,\uparrow\downarrow\rangle)$ & 
 0 &  1.36E-4 & 1.36E-4 &  -1.36E-4  &
 0 & -2.66E-5 & -2.66E-5 &  2.66E-5  \\
${1\over \sqrt{2}} (|\uparrow\downarrow,0\rangle-|0,\uparrow\downarrow\rangle)$ & 
${1\over \sqrt{2}} (|\uparrow\downarrow,\uparrow\rangle-|\uparrow,\uparrow\downarrow\rangle)$ & 
4.52E-6 & 0 & 4.52E-6 &  4.52E-6  &
-3.48E-7 & 0 & -3.48E-7 &  -3.48E-7  \\
${1\over \sqrt{2}} (|\uparrow,\downarrow\rangle+|\downarrow,\uparrow\rangle)$ & 
${1\over \sqrt{2}} (|\uparrow\downarrow,\uparrow\rangle-|\uparrow,\uparrow\downarrow\rangle)$ & 
0 & 6.80E-5 & 6.80E-5 &  -6.80E-5  &
0 & -1.33E-5 &  -1.33E-5 &   1.33E-5  \\ 
${1\over \sqrt{2}} (|\uparrow\downarrow,\downarrow\rangle+|\downarrow,\uparrow\downarrow\rangle)$ & 
$|\uparrow\downarrow,\uparrow\downarrow\rangle$ & 
0 & 9.04E-6 & 9.04E-6 &  -9.04E-6  &
0 & 7.05E-7 & 7.05E-7 &  -7.05E-7  \\
\hline
$\langle n|A^\dag|G\rangle$ (sum) && 
2.99E-4&2.99E-4 & 5.98E-4 & $\sim$0  &
-2.39E-5 &2.43E-5 & $\sim$0 & -4.82E-5  \\
\hline
\end{tabular}
\caption{$M_{\alpha',\alpha}(A)$ between the ground state and a  NRG excitation state for $I=0.002$ (Kondo resonance regime) and $I=0.003$(RKKY dominated regime).}
\label{tab:matrixelement}
\end{table*}

\section{Resonance-pseudogap ``duality'' in the particle-hole asymmetric case}
\label{sec:phasym}

In Sec.~\ref{sec:cfermion}, we have shown that in the particle-hole symmetric case, there exists a resonance-pseudogap ``duality'' in the spectral functions of the regular fermion operator $f_{i\sigma}$ and the composite fermion operator ${\bar f}_{i\sigma} = f_{i\sigma}(1-2n_{i{\bar \sigma}})$, i.e., the regular fermion has a resonance-shape spectral function while the composite fermion has a pseudogap-type spectral function in the Kondo resonance regime, and vice versa in the RKKY-dominated regime.  In the particle-hole asymmetric case, e.g., when $\epsilon_f \neq -U/2$,  the average occupancy number of Anderson orbitals is no longer 1. For a paramagnetic solution, $\langle n_{i\uparrow} \rangle =\langle n_{i\downarrow}\rangle = \langle n_i \rangle/2 \neq 1/2$, and $f_{i\sigma}$ and ${\bar f}_{i\sigma}$ are no longer orthogonal to each other $\langle [f_{i\sigma},  {\bar f}_{i\sigma}^\dag]_+\rangle = \langle (1-2n_{i{\bar \sigma}})\rangle \neq 0$. In practice, we find that both types of fermions have finite spectral weights at the Fermi energy in either regimes.  We show below that the resonance-pseudogap ``duality'' can be recovered by two new orthogonal modes after a transformation.  

We take as an example the two-impurity Anderson model with $U=2$ but $\epsilon_f \neq -U/2$. Similar to the first case in the main text, we choose the hybridization functions in the even and odd parity channels to be the same $\Gamma_e = \Gamma_o = 0.045\pi$ for $|\omega| \leq 1$, and use an explicit spin-exchange term $I{\bf S}_1\cdot {\bf S}_2$ to tune from the Kondo resonance regime to the RKKY dominated regime.  Since no parity symmetry breaking term is present, it is found that a quantum critical point (QCP) still exists associated with the divergence of the staggered spin susceptibility ${\bf S}_1-{\bf S}_2$. However, both the uniform charge susceptibility $\chi_{u,ch}=\langle\langle n_1+n_2; n_1+n_2\rangle\rangle$ and the staggered charge susceptibility $\chi_{a,ch}=\langle\langle n_1-n_2; n_1-n_2\rangle\rangle$ are found to diverge, in contrast to the QCP in the particle-hole symmetric case $\epsilon_f=-U/2$, where only $\chi_{a,ch}$ diverges. 
 
\begin{figure}[tbh]
\centering
\includegraphics[width=\columnwidth]{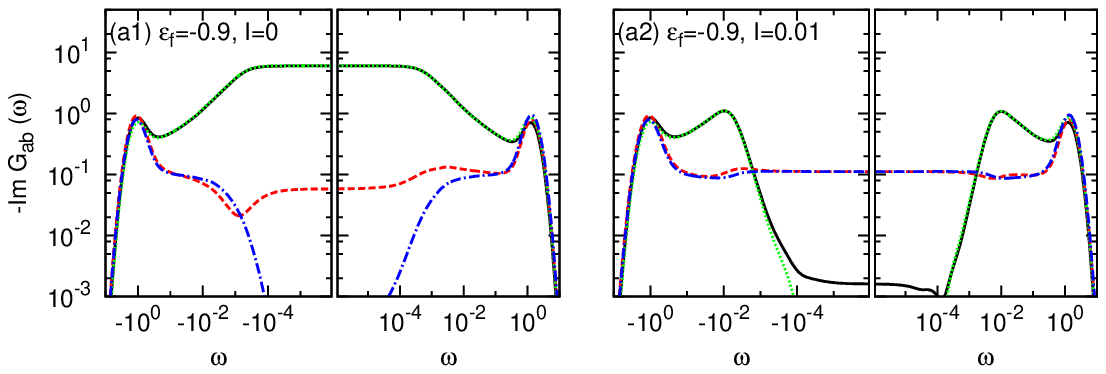}
\includegraphics[width=\columnwidth]{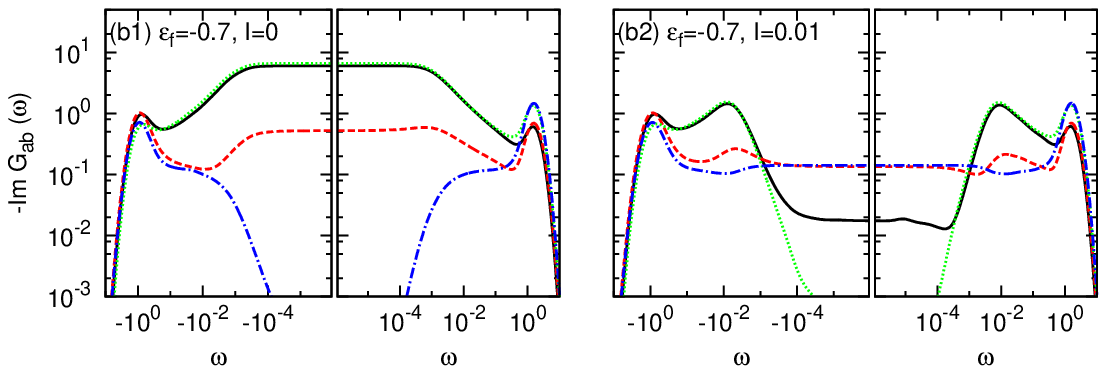}
\includegraphics[width=\columnwidth]{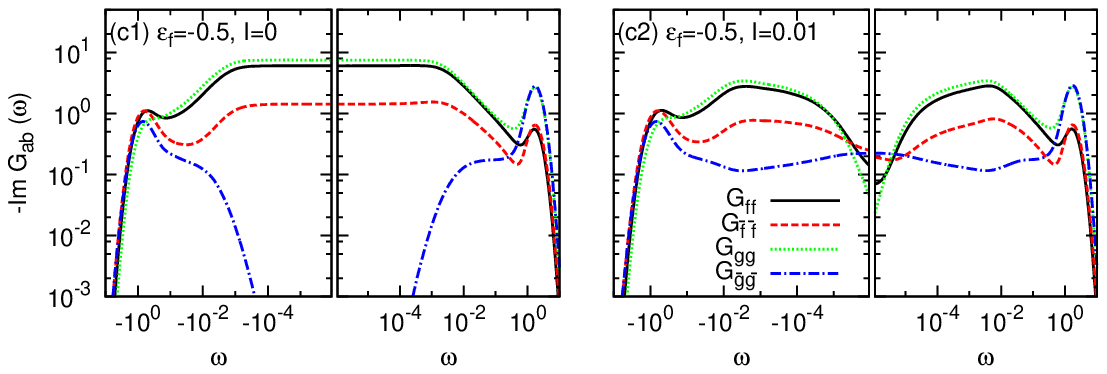}
\caption{ (Color online) The spectral functions in the particle-hole asymmetric case. Different color lines represent $\text{-Im}G_{ff}$ (black), $\text{-Im}G_{{\bar f}{\bar f}}$ (red),  $\text{-Im}G_{gg}$ (green), and $\text{-Im}G_{{\bar g}{\bar g}}$ (blue). (a)-(c) are with different values of $\epsilon_f = -0.9, -0.7,-0.5$, away from the particle-hole symmetry $\epsilon_f =-1$. Other parameters are $\Gamma_{e,o}(\omega)=0.045\pi$ for $|\omega|<1$, $U=2$ and $\mu=0$. For each value of $\epsilon_f$, we show results of two values of $I=0,0.01$, in (1), (2), as representative cases in the Kondo and RKKY dominated regimes. }
\label{fig:cf-asym}
\end{figure}

In Fig.~\ref{fig:cf-asym}, we show the results of the spectral functions ($-\text{Im}G$) of $f_{p\sigma}$ and ${\bar f}_{p\sigma}$ for $\epsilon_f$ =-0.9, -0.7, and -0.5. For each case, we choose $I=0$ and $I=0.01$ to represent the Kondo and the RKKY dominated regimes, respectively. It can be found that the spectra of $f_{p\sigma}$ changes from the Kondo resonance form ($I=0$) to a pseudogap form ($I=0.01$). However, its spectral weight at $\omega=0$, although small, is always finite. Conversely, the spectra of ${\bar f}_{p\sigma}$ have a finite spectral weight in the Kondo resonance regime.  In practice, we find that $\text{Im}G_{ee}(0) \text{Im}G_{dd}(0) = [\text{Im}G_{de}(0)]^2$ is always satisfied (we do not have an exact proof of this relation yet; presumably, it is due to a certain sum rule associated with $[d_{i\sigma}, e^\dag_{i\sigma}]_+=0$). Here, $-\text{Im}G_{de}(0)$ changes sign from the Kondo resonance regime (positive) to the RKKY dominated regime (negative). Therefore, $-\text{Im}G_{ff}(0) \text{ or } [-\text{Im}G_{{\bar f}{\bar f}}(0)]$ can be expressed as  $(\sqrt{-\text{Im}G_{ee}(0)}\pm \sqrt{-\text{Im}G_{dd}(0)})^2$ in the Kondo resonance regime and vice versa in the RKKY dominated regime. In the particle-hole asymmetric cases, due to $\text{Im}G_{ee}(0) \neq \text{Im}G_{dd}(0)$ (the former has a larger spectral weight when $\langle n \rangle <1$), $-\text{Im}G_{ff}(0)$ and $-\text{Im}G_{{\bar f}{\bar f}}(0)$ are always finite.

\begin{figure}[tbh]
\centering
\includegraphics[width=0.8\columnwidth]{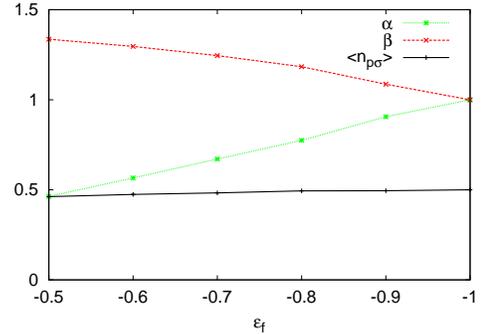}
\caption{(Color online) Parameters of ($\alpha$, $\beta$) in the transformation $g_{p\sigma} = \alpha d_{p\sigma} + \beta e_{p\sigma}$ and ${\bar g}_{p\sigma} =  \alpha d_{p\sigma} - \beta e_{p\sigma}$ for different values of $\epsilon_f$. Also shown is the particle number $\langle n_{p\sigma}\rangle= - (1/\pi) \int _{-\infty}^0 \text{Im} G_{ff,p\sigma}(\omega)$.}
\label{fig:param}
\end{figure}

This motivates us to introduce a transformation 
\begin{eqnarray}
g_{p\sigma} &=& \alpha e_{p\sigma} + \beta d_{p\sigma}, \nonumber \\
{\bar g}_{p\sigma} &=&  \alpha e_{p\sigma} - \beta d_{p\sigma}, 
\label{eq:transform}
\end{eqnarray}
which is a SO(2) rotation in ($e_{p\sigma}$, $d_{p\sigma}$) space,
such that $\alpha^2 \text{Im}G_{ee}(0) =\beta^2  \text{Im}G_{dd}(0)$. It follows that $g_{p\sigma}={\alpha+\beta \over 2} f_{p\sigma} + {\alpha-\beta \over 2} {\bar f}_{p\sigma}$ and ${\bar g}_{p\sigma}={\alpha-\beta \over 2} f_{p\sigma} + {\alpha+\beta \over 2} {\bar f}_{p\sigma}$.
The spectral functions of  $g_{p\sigma}$ and ${\bar g}_{p\sigma}$ are also shown in Fig.~\ref{fig:cf-asym}. For different values of $I$ with the same $\epsilon_f$, we use only one set of parameters ($\alpha$,$\beta$).  After the transformation, we observe that the resonance-pseudogap ``duality'' is established between $g_{p\sigma}$ and ${\bar g}_{p\sigma}$. Besides, their spectral functions are more particle-hole symmetric near the Fermi energy. 
The results of ($\alpha$,$\beta$)  for different values of $\epsilon_f$ are shown in Fig.~\ref{fig:param}. As $\epsilon_f \to -1$, the particle-hole symmetry case, $g$ and ${\bar g}$ become $f$ and ${\bar f}$. 

Such a transformation can also explain the divergence of the uniform charge susceptibility. We find that in the new basis, the charge susceptibility $\langle\langle n_{g1}-n_{g2}; n_{g1}-n_{g2} \rangle\rangle$ is divergent, while $\langle\langle n_{g1}+n_{g2}; n_{g1}+n_{g2} \rangle\rangle$ is not at the QCP.  As $n_{g1}-n_{g2}$ has a finite projection to $n_1 +n_2$ in the particle-hole asymmetric case, as well as the finite projection to $n_1-n_2$. Therefore, both the uniform and staggered charge susceptibilities are expected to diverge.  The similar phenomenon has also been observed in the two-impurity Anderson model with a finite magnetic field, where both the uniform and staggered spin susceptibilities are found to be divergent at a field-tuned quantum critical point.~\cite{zhu2011} Here, the chemical potential $\mu$, which couples the total particle number, acts as the ``magnetic field'' in pseudospin space. 

\section{Cooper pair correlations}
\label{sec:pair}

\begin{figure}[tbh]
\centering
\includegraphics[width=\columnwidth]{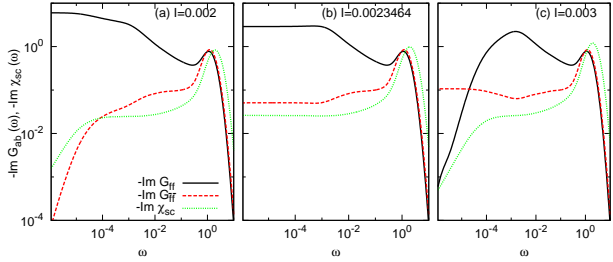}
\caption{(Color online) The imaginary part of the Cooper pair correlation function $\chi_{sc}$ (green), the single-particle Green's function $G_{ff}$ (black), and the Green's function $G_{{\bar f}{\bar f}}$ for the composite fermion  (red) for three representing values of RKKY interaction in $I<I_c$, $I\approx I_c$, and $I>I_c$. }
\label{fig:sc}
\end{figure}

We study the Cooper pair correlation functions $\chi_{sc} = \langle \langle \Delta_{sc}; \Delta_{sc}^\dag \rangle \rangle$, where $\Delta_{sc}^\dag = f^\dag_{1\uparrow} f^\dag_{2\downarrow} + f^\dag_{2\uparrow}f^\dag_{1\downarrow}=J_e^+-J_o^+$ and $J^+_p =f^\dag_{p\uparrow} f^\dag_{p\downarrow}$. Such an intersite singlet pair is favored over the onsite one $ f^\dag_{1\uparrow} f^\dag_{1\downarrow} + f^\dag_{2\uparrow}f^\dag_{2\downarrow}=J_e^++J_o^+$ due to repulsive interaction $U$, as written in parity basis, $\sum_i Un_{i\uparrow}n_{i\downarrow}$ contains a term $U J_e^+J_o^-$, which acts as an ``antiferromagnetic'' coupling between the pseudospins ${\bf J}_e$ and ${\bf J}_o$.~\cite{ZhuSC} 
 
In Fig.~\ref{fig:sc}, we show the results of $-\text{Im} \chi_{sc} (\omega)$ together with the single-particle spectra for $f^\dag_{p\sigma}$ and ${\bar f}^\dag_{p\sigma}$. We notice that the finite pair excitations are closely related to the presence of both finite excitations of $f^\dag_{p\sigma}$ and ${\bar f}^\dag_{p\sigma}$. At low energies, this only takes place in the vicinity of quantum critical point [Fig.~\ref{fig:sc}(b)]. Here, the spectral weights for $f^\dag_{p\sigma}$ and ${\bar f}^\dag_{p\sigma}$ are half of their corresponding ones in the Kondo resonance regime  [Fig.~\ref{fig:sc}(a)] and the RKKY dominated regime [Fig.~\ref{fig:sc}(c)], respectively. 

As ${\bar f}_{e\sigma} = f_{e\sigma}(n_{e{\bar \sigma}}+n_{o{\bar \sigma}}-1) + f^\dag_{e{\bar \sigma}}f_{o{\bar\sigma}}f_{o\sigma}-f_{e{\bar \sigma}}f^\dag_{o{\bar \sigma}}f_{o{\sigma}}$, it can be argued that an enhanced ``hybridization'' term $f_{e\sigma}^\dag{\bar f}_{e\sigma}$ also promotes the pairing term $f^\dag_{e\sigma}f^\dag_{e{\bar \sigma}}f_{o{\bar\sigma}}f_{o\sigma} \sim \Delta_{sc}^\dag\Delta_{sc}$, as well as other instabilities. These interaction parameters have also been fitted from the leading irrelevant parameters in the low-energy effective Hamiltonian, which are found to be divergent (gapless) at the QCP.~\cite{Jones87}


\begin{thebibliography}{99}
\bibitem{footnote-se} $H_{\text{int}}$ also contains the added RKKY term $I{\bf S}_1 \cdot {\bf S}_2$. But it is found the latter has negligible contributions to the self-energy.
\bibitem{footnote} The current numerical calculation cannot capture accurately the low energy form of $-\text{Im}\Sigma_{p\sigma}$. From $\text{Re}\Sigma_{p\sigma}(\omega) \sim 1/\omega$, we expect $-\text{Im}\Sigma_{p\sigma}(\omega) \sim \delta(\omega)$.
\end{thebibliography}
\end{document}